\documentclass[pra,aps,twocolumn,floatfix]{revtex4-1}
\usepackage{graphics,color}
\usepackage{amsmath}
\usepackage{changebar}
\usepackage{footmisc}
\usepackage{epsfig}
\usepackage{amsfonts}
\usepackage{amssymb}
\usepackage{amsbsy}
\usepackage{xcolor}
\usepackage{braket}

\begin{document}

\title{Ordering kinetics in $q$-state clock model: scaling properties and growth laws}
\author{Swarnajit Chatterjee$^1$, Sanjay Puri$^2$\footnote{author for correspondence: ssprp@iacs.res.in} and Raja Paul$^1$\footnote{author for correspondence: purijnu@gmail.com}}

\affiliation{$^1$ Indian Association for the Cultivation of Science, Kolkata -- 700032, India.}
\affiliation{$^2$ School of Physical Sciences, Jawaharlal Nehru University, New Delhi -- 110067, India.}

\begin{abstract}
We present a comprehensive Monte Carlo study of the ordering kinetics in the $d=2$ ferromagnetic $q$-state clock model with nonconserved Glauber dynamics. In agreement with previous studies we find that $q \geqslant 5$ is characterized by two phase transitions occuring at temperatures $T_{c}^1$ and $T_{c}^2$ ($T_{c}^2<T_{c}^1$). Phase ordering kinetics is then investigated by rapidly quenching the system in two phases, in the quasi-long range ordered phase (QLRO) where $T_{c}^2<T<T_{c}^1$ and in the long-range ordered phase (LRO) where $T<T_{c}^2$; $T$ being the quench temperature. Our numerical data for equal time spatial correlation function $C(\textbf{r},t)$ and structure factor $S(k,t)$ support dynamical scaling. Quench in the LRO regime is characterized by a crossover from an preasymptotic growth driven by the annealing of both vortices and interfaces to an interface driven growth at the asymptotic regime with growth exponent $n\simeq 0.5$. In the QLRO quench regime, domains coarsen mainly via annihilation of point defects and our length scale data for $q$ = 9, 12, and 20 suggests a $R(t) \sim (t/\ln t)^{1/2}$ growth law for the $q$-state clock model in the QLRO phase. 
\end{abstract}

\maketitle
\section{Introduction}
In statistical physics, the $q$-state clock model is considered as the discrete version of the generalized $XY$ model. Theoretical interest in clock model was stimulated after Kosterlitz and Thouless (KT) in their pioneering works \cite{KT,KT74} showed that the $XY$ model possessed a novel type of critical behavior with essential singularities and topological ordering. The $q$-state clock model ground state is $q$-fold degenerate where the clock spins can take only discrete angles of the continuous $XY$ spins from a set of values governed by the $q$ value and the continuous $U(1)$ symmetry is replaced by the discrete $Z_q$ symmetry. This is essentially equivalent to probing the $q$-fold symmetry-breaking fields and the effect of this $q$-fold symmetry-breaking fields on the $d=2$ $XY$ model has been the subject of attention for many years\cite{kadanoff77,elitzur79,domany80,cardy80,tobochnik82,baek2010,brito2010}. 

The $q$-state clock model shows Ising like phase transitions for $q$ $\leqslant$ $4$ and two distinct phase transitions at finite temperatures $T_{c}^1$ and $T_{c}^2$ ($<T_{c}^1$) for $q \geqslant 5$ \cite{kadanoff77,elitzur79,domany80,cardy80,tobochnik82,baek2010,brito2010}. The phase between these temperatures are quasi-long range ordered (QLRO) like the $XY$ model below the KT temperature;  the phase above $T_{c}^1$ is high temperature disordered phase and the phase below $T_{c}^2$ is long-range ordered (LRO) \cite{papa2012}. There are studies \cite{rujan81,roomany81} which suggest that for $q$ = 5, the number of phase transitions are inconclusive, but it has been shown extensively in literatures that for planer 5-state clock model, there happens two transitions where the upper transition is KT like \cite{domany80,cardy80,papa2011}. For $q$ = 6, there exists some controversies regarding the KT like nature of the transition at $T_{c}^1$ \cite{wexler2006,hwang2009}, but it has been comprehensively established over the years that the transitions occurring at $T_{c}^1$ and $T_{c}^2$ for $q>4$ are indeed of KT type \cite{kadanoff77,brito2010,miyashita78,landau86,ono91,okabe2002,tomita2002,tomita2001,rastelli2004,surungan2005,
kim2010,wu2012}. The overall effect of state $q$ on $T_{c}^1$ and $T_{c}^2$ suggests that while $T_{c}^1$ does not change appreciably for large $q$ and tends to merge with the KT transition temperature $T_{KT} \sim 0.89$, $T_{c}^2$ keeps decreasing with the increasing $q$ \cite{brito2010,wexler2006}. 

Systematic characterization of the phase transition points of the $q$-state clock model and ordering kinetics for an extended set of $q$ in both QLRO and LRO regimes is the primary focus of this study. Phase ordering kinetics of various systems quenched from a high-temperature disordered phase to a low-temperature ordered phase has been studied widely to investigate the domain growth law and the dynamical scaling behavior of the correlation function and structure factor \cite{Bray94}. The characteristic length scale $R(t)$ typically grows as $R(t)\sim t^n$, where $n$ is the `growth exponent'. The domain growth law varies depending upon whether the order parameter is conserved or not. For the $q$-state clock model with conserved order parameter it has been shown that a slow domain growth in the early time-regime consistent with the growth law for the corresponding $XY$ model, crosses over to a faster growth at the asymptotic limit consistent with the Lifshitz-Slyozov growth law $R(t) \sim t^{1/3}$ \cite{puri97}. In this study, we consider clock model with non-conserved order parameter, which in the asymptotic limit follows the Lifshitz-Cahn-Allen (LCA) growth law: $R(t) \sim t^\frac{1}{2}$ \cite{Bray94,PW}. The clock model is highly significant as it interposes between the Ising model ($q=2$) and the $XY$ model ($q= \infty$). Coarsening in the Ising model is driven by the merging of interfaces, whereas annihilations of vortices and antivortices govern the domain growth in the $XY$ model. Interestingly, in the $q$-state clock model, coarsening occurs via the elimination of both interfaces and vortices. Literature suggests that nonequilibrium kinetics and scaling of the correlation function of the $q$-state clock model \cite{KG83,KG85,KNG85,GS84,EK89,EK90,corberi2006} and Potts model \cite{Kaski87} in the LRO regime marked by the LCA growth law. Analytical studies on these systems also confirm dynamical scaling of the correlation function and structure factor and suggest that the latter is a function of $q$ \cite{LM93,kawasaki85,SM95}. Coarsening dynamics of the $q$-state clock model following a quench in the QLRO phase has not found much attention as literature points to the study by Corberi $et.$ $al.$ \cite{corberi2006} where the authors have briefly mentioned the QLRO domain growth process for 6-state clock model. 

Here, we present a study of transition temperatures $T_{c}^1$ and $T_{c}^2$ for various $q$ values via Wolff single-cluster update algorithm \cite{wolff89}. This enables us to figure out the regime for temperature quench in both the LRO and QLRO phases. Subsequently, we study the ordering kinetics in the $q$-state clock model using Metropolis algorithm \cite{metropolis53} following a temperature quench in both LRO and QLRO regime. The main results of our study are summarized below:

(a) For $q \geqslant 5$, $T_{c}^1$ remains almost independent of $q$, whereas, $T_{c}^2$ decreases with $q$.

(b) Coarsening dynamics following a quench from $T$ = $\infty$ to $T<T_{c}^2$ (LRO regime) is characterized by the curvature driven domain growth law $R(t) \sim t^{1/2}$ at the asymptotic limit.

(c) Interpenetrating domains with rough domain interfaces are typical of the quench from $T$ = $\infty$ to $T_{c}^2<T<T_{c}^1$ (QLRO regime). The system exhibits slow domain growth and the growth law we extract over our simulation time-scales is $R(t) \sim (t/\ln t) ^{1/2}$ for higher values of $q$.

The paper is organized as follows. In Sec.~\ref{Modeling}, we discuss the model and present detailed description of numerical simulations scheme. In Sec.~\ref{Results}, we present detailed numerical simulation results for $d=2$ clock model. Finally, in Sec.~\ref{Summary}, we conclude this paper with a summary and discussion of the results. 

\section{Modeling and Simulation Details} 
\label{Modeling}

\subsection{$q$-state Clock model}
The Hamiltonian for the $q$-state clock model is defined as

\begin{equation}
\label{refH}
\mathcal{H} = -J\sum\limits_{\langle ij \rangle} \vec{\sigma_i} \cdot \vec{\sigma_j} = -J\sum\limits_{\langle ij \rangle} \cos ({\theta}_i - {\theta}_j) ,
\end{equation} 
where $\langle ij \rangle$ denotes nearest neighbor sites. In Eq.~\eqref{refH}, $\vec{\sigma_i}$ denotes a two-component unit vector spin; e.g., in xy plane $\vec{\sigma_i}$ = $\hat{x} \cos \theta_i+\hat{y} \sin \theta_i$. The unit vector $\vec{\sigma_i}$ is described by an angle $\theta_i$ $\in$ (0, 2$\pi$) where
\begin{equation}
\label{theta}
\theta_i = \frac{2\pi n_i}{q},
\end{equation}
and $n_i$ = 0, 1, 2, ...., ($q-1$) denote discrete orientations of the spin. $J$ is the coupling between neighboring sites and is taken as 1.

\subsection{Simulation details for study of transition temperatures}

In our study, we first revisit the well known problem of equilibrium phase transition in the $q$-state clock model to precisely identify the regime for temperature quench. In our simulations, canonical sampling Monte Carlo (MC) method with Wolff single cluster flipping algorithm \cite{wolff89} is applied to equilibrate the system during the characterization of equilibrium thermodynamic parameters. A single Monte Carlo step (MCS) update is described as$\colon$ 

(a) A random reflection with a normal vector $\vec{r}$ = $(\cos {\delta}_i,\sin {\delta}_i)$ and a random spin $\vec{\sigma_i}=(\cos \theta_i,\sin \theta_i)$ are chosen as starting points for the cluster $\mathcal{C}$. 

(b) The spin is given a reflection $\mathcal{R}(\vec{r})\vec{\sigma_i}$ = $\vec{\sigma_i}-2(\vec{\sigma_i} \cdot \vec{r})\vec{r}$ about the line; $i.e.$ $\theta_i\rightarrow \theta_i^{\prime}=\pi-\theta_i+2\delta_i$, where $\theta_i$ is the primary angle of the site $i$ and $\theta_i^{\prime}$ is the angle after reflection and $\delta_i$ = $i(\frac{\pi}{q})$ for even $q$ and $\delta_i$ = $(i+\frac{1}{2})(\frac{\pi}{q})$ for odd $q$ with $i$ = 0, 1, 2, ...., ($2q-1$) \cite{wu2012}.

(c) The reflected position of the spin is again marked and nearest neighbors of the spin are visited and if the spins do not belong to the cluster they are added to the cluster according to a probability $\mathcal{P}(\vec{\sigma_i},\vec{\sigma_j})$ = $1-\exp (min [0, 2 \beta J_{ij}(\vec{r} \cdot \vec{\sigma_i})(\vec{r} \cdot \vec{\sigma_j})])$ or $\mathcal{P}(\theta,\delta)$ = $\cos (\theta_i-\delta)\cos (\theta_j-\delta)$ \cite{wolff89}. 

Finally, the cluster is updated by reflecting all the spins about the line perpendicular to the normal vector $\vec{r}$. If $N$ denotes total number of sites, then one MCS corresponds to $N$ such updates.

Measurements of the thermodynamic parameters is carried out after the system has reached thermal equilibrium. We measure the magnetic order parameter $m$, defined by the equation 
\begin{equation}
\label{mag}
m=\frac{1}{N} \sqrt{\left(\sum_{i=1}^N\cos\theta_i\right)^2+\left(\sum_{i=1}^N\sin\theta_i\right)^2},
\end{equation}
and per spin specific heat $C_v$ defined as 
\begin{equation}
\label{cv}
C_v=\frac{1}{Nk_BT^2}[\langle E^2 \rangle-\langle E \rangle^2], 
\end{equation}
where $T$ is the temperature, $k_B$ is the Boltzmann constant ($k_B$ = 1) and $E$ is the total energy per spin defined in Eq.~\eqref{refH}. 

The Binder cumulant $U_4(T, L)$ \cite{binder2005,NB,binder81,loison99} expressed as
\begin{equation}
\label{BC}
U_4= 1-\frac{\langle m^4 \rangle}{3\langle m^2 \rangle ^2}, 
\end{equation}
and plotted against $T$, can precisely quantify the transition temperature from the intersection of the curves for various $L$. This mechanism has been implemented to determine the upper critical temperature $T_{c}^1$. Nevertheless, $U_4$ could not detect the transition between the QLRO and LRO phase due to $m$ constructed using square of the sum of the spin components which can not distinguish between the orientations of the spin vectors in these two phases. Following \cite{baek2009}, we define the resultant angular direction of the spins, $\phi$ = $\tan^{-1}{\big(\frac{\sigma_y}{\sigma_x}\big)}$, where $\sigma_x$ = $\sum_{i=1}^N\cos\theta_i$ and $\sigma_y$ = $\sum_{i=1}^N\sin\theta_i$. We now define an effective order parameter $m_\phi =\langle \cos(q\phi) \rangle$ and a cumulant $U_m$$\colon$
\begin{equation}
\label{um}
U_m=1-\frac{\langle m_\phi^4 \rangle}{2\langle m_\phi^2 \rangle ^2}, 
\end{equation} 
In the same spirit as $U_4$, $U_m$ plotted against $T$ can quantify $T_{c}^2$. $U_m$, however, is not a suitable cumulant to measure $T_{c}^1$ as in the high temperature homogeneous phase $\phi$ might become undefined.

\subsection{Simulation details for study of ordering kinetics}
At the outset for studying ordering kinetics in the $q$-state clock model, we assign random initial orientation to each spin $\theta_i$, defined in Eq.~\eqref{theta}, to mimic the high temperature disordered phase. Followed by, we rapidly quench the system independently at temperature $T$ into two regimes, 1. $T_{c2}<T<T_{c1}$ and 2. $T<T_{c2}$ at $t$ = 0 and let the system evolve via nonconserved Glauber kinetics up to $t$ = $10^6$ MCS using Metropolis algorithm \cite{metropolis53}. The algorithm is the following:

(a) A random spin $\vec{\sigma_i}$ is chosen and $\theta_i$ is given a small rotation $\delta_i$ $\in$ $\frac{2\pi s_i}{q}$, $s_i$ = 1,..., $q-1$.
(b) The new spin state $\theta_i^{\prime}$ = $\theta_i + \delta_i$ is accepted with the probability $P$ = $\text{min}[1,\text{exp}(-\beta\Delta\mathcal{H})]$, where $\Delta\mathcal{H}$ is the change in energy resulting from spin change $\theta_i\rightarrow \theta_i^{\prime}$ and can be expressed as:
\begin{equation}
\Delta\mathcal{H}=\sum_{k}J_{ik}\Big\{\cos(\theta_i-\theta_{k})-\cos(\theta_i^{\prime}-\theta_{k})\Big\},
\end{equation}
where $k$ refers to the nearest neighbors of site $i$.

The segregation kinetics of the $q$-state clock model can be investigated by studying the time dependence of the correlation function $C(\textbf{r},t)$ expressed as \cite{PW}:
\begin{eqnarray}
\label{refC}
C({\vec{r}},t)&=&\frac{1}{N}\sum_{i=1}^{N} [\langle \vec{\sigma_i}(t) \cdot \vec{\sigma_{i+\textbf{r}}}(t) \rangle - \langle \vec{\sigma_i}(t) \rangle \cdot \langle \vec{\sigma_{i+\textbf{r}}}(t) \rangle]_{av}\nonumber \\
&=&\frac{1}{N}\sum_{i=1}^{N} [\langle \cos\big\{\theta_i(t)-\theta_{i+\textbf{r}}(t)\big\} \rangle]_{av},
\end{eqnarray} 
where $[\langle ... \rangle]_{av}$ indicates an average over different initial realizations. Another commonly used probe for domain growth is the structure factor, which is defined as the Fourier transform of the correlation function \cite{PW},
\begin{equation}
\label{SF}
S({\vec{k}},t)=\int d\vec{r} e^{i\vec{k}\cdot\vec{r}}C({\vec{r}},t),
\end{equation}
where $\vec{k}$ is the wave vector of the scattered beam. Scattering experiments measure the structure factor $S(\vec{k},t)$. Isotropically, $C({\vec{r}},t)$ and $S({\vec{k}},t)$ depend upon the absolute value of the vectors, $r$ = $|\vec{r}|$ and $k$ = $|\vec{k}|$, respectively.

If the system is isotropic and characterized by a single length scale $R(t)$, domain morphologies does not change with time $t$ apart from a scale factor. The correlation function and structure factor exhibit the following dynamical scaling forms \cite{PW}:
\begin{equation}
\label{CFScaling}
C({\vec{r}},t) = f\left(\frac{r}{R}\right),
\end{equation} 

\begin{equation}
\label{SFScaling}
S(\vec{k},t)=R(t)^dg[kR(t)],
\end{equation}
where $d=2$ refers to the dimensionality. The scaling functions $f(x)$ and $g(y)$ are related as
\begin{equation}
g(y)=\int d\vec{x} e^{i\vec{x}\cdot\vec{y}}f(x),
\end{equation} 
The characteristic length scale $R(t)$ is defined as the distance over which the correlation function $C({\vec{r}},t)$ decays to an arbitrary fraction (e.g. 0.3) of its maximum value. Asymptotically, the only existing characteristic length scale is $R(t)$ \cite{Bray94} which could be extracted either from the decay of $C({\vec{r}},t)$ or from the number density of defects. In this paper, $R(t)$ is determined from the decay of $C({\vec{r}},t)$. The morphology of domain structure and the coarsening dynamics can also be viewed from the analysis of structure factor $S(\vec{k},t)$. Bray, Puri \cite{BP91}, and Toyoki \cite{toyoki92} have independently proposed that for a $n$-component vector field, the scaling function $g(y)$ has a large-$y$ behavior: 
\begin{equation}
\label{porod}
g(y)\sim y^{-(d+n)},\quad \text{for} ~~ kR\rightarrow \infty.
\end{equation} 
Eq.~\eqref{porod} is known as the \textit{generalized Porod tail} reduces to the famous Porod's law in scaler order parameter field $n=1$ \cite{POROD82,OP88} and recognized as emerging from the configurations of sharp defect-interfaces.


\section{Numerical Results} 
\label{Results}

In this section, we present numerical results from our simulations of the two-dimensional $q$-state clock model. Initially, we have estimated the transition temperatures ($T_{c}^1$ and $T_{c}^2$) for various spin states $q$ marking the phase diagram relevant to the interest of this paper. Subsequently, we study the coarsening dynamics in the $q$-state clock model following two independent temperature quench in the LRO regime ($T<T_{c}^2$) and QLRO regime ($T_{c}^2<T<T_{c}^1$). 

\subsection{Estimation of $T_{c}^1$ and $T_{c}^2$}

Let us first present the results which quantify $T_{c}^1$ and $T_{c}^2$. We study the $q$-state clock model on a square lattice ($L^2$) of linear sizes $L$ = 32, 64, 96, 128 and 256. Starting from a random initial configuration, the system has been equilibrated using Wolff cluster update algorithm~\cite{wolff89}. After equilibrating the system for $\sim 10^6$ MCS, we thermally average $m$, $C_v$, $m^2$, $m^4$, and $m_{\phi}$ upto $6 \times 10^5$ MCS. One Monte Carlo step (MCS) corresponds to attempted sweep across the whole lattice ($L^2$). The results are further averaged over 100 independent runs of initial configurations. Our results confirm that for $q$ $\geqslant$ 5, two phase transitions occur: one at a low-temperature ($T_{c}^2$) and the other at a relatively higher temperature ($T_{c}^1$) which concur with earlier findings \cite{kadanoff77,elitzur79,domany80,cardy80,tobochnik82,baek2010,brito2010}.

Although existence of these transitions can be visualized from the magnetization $m$ and the peaks of the specific heat $C_v$ plotted against $T$, precise quantification of the transition temperatures would require extensive simulation with system size $L\rightarrow \infty$. Thus, fourth order cumulant of the relevant order parameter \cite{binder2005,NB,binder81,loison99} $U_4$ used as a preferred method of estimating the upper transition temperature $T_{c}^1$; nevertheless, $U_4$ failed to capture the lower transition temperature $T_{c}^2$ for $q$ $>$ 4. Therefore, $T_{c}^2$ for different $q$'s have been measured from the temperature dependency of the cumulant $U_m$ defined in Eq.~\eqref{um}. 

\begin{figure}[!htbp]
\centering
\includegraphics[width=\columnwidth]{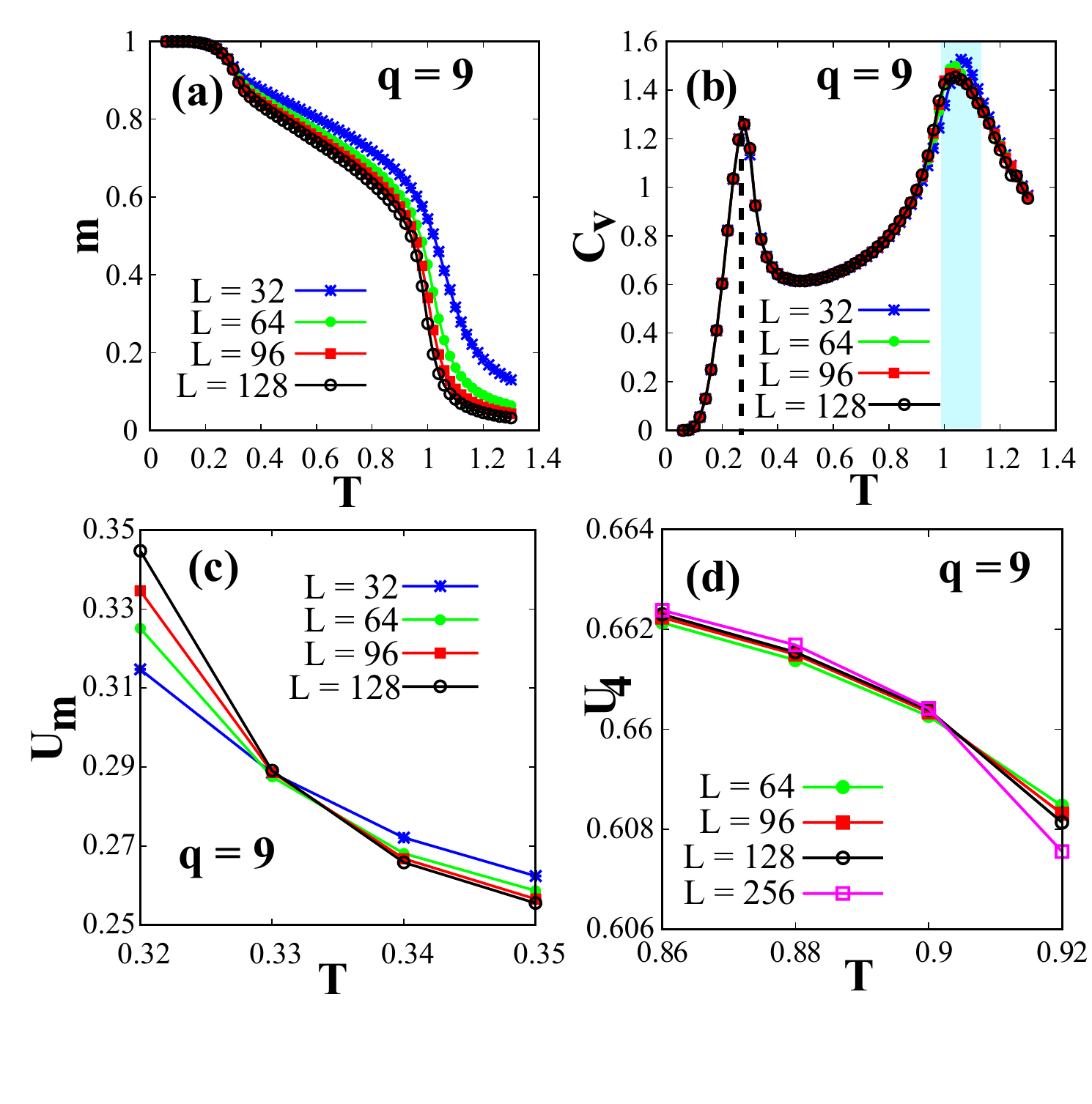}
\caption{(Color online) Equilibrium parameters of the 9-state clock model for $L$ = 32 (blue star), 64 (green solid circle), 96 (red solid square), 128 (black open circle) and 256 (magenta open square). (a) Magnetization $m$ versus $T$ for 9-state clock model. Two inflections observed in the profile of $m$ correspond to the phase transitions. (b) Two peaks in the $C_v$ versus $T$ plot confirms the fact of two different phase transitions; right peak denotes transition from disordered to QLRO phase occurring at higher temperature $T_{c}^1$ and transition from QLRO to ordered phase is denoted by the left peak at a lower temperature $T_{c}^2$. The shaded region in (b) implies the spread of the peak for which an accurate measure of $T_{c}^1$ is not possible. A more precise quantification of $T_{c}^2$ and $T_{c}^1$ from the temperature dependence of $U_m$ and $U_4$ are shown in (c) and (d) respectively.}
\label{fig1}
\end{figure}

Fig.~\ref{fig1} shows data for the equilibrium properties of the 9-state clock model for $L$ = 32, 64, 96, 128 and 256. In the figure the linear lattice lengths are respectively represented by blue star, green solid circle, red solid square, black open circle and magenta open square (color online). Two distinct regions of inflection in Fig.~\ref{fig1}(a) correspond to two different transitions: one from disordered to quasi-long range ordered phase via $T_{c}^1$ and another from quasi-long range ordered phase to the ordered phase via $T_{c}^2$. Two distinct peaks in $C_v$ versus $T$ plot in Fig.~\ref{fig1}(b) confirms this scenario where the right peak signifies a phase transition from the disordered homogeneous phase to the QLRO phase and the left peak defines the phase transition from QLRO to LRO (ordered) phase. Notice, that the right peak which corresponds to the upper critical temperature $T_{c}^1$, decreases as $L$ increases, but the change is not significant for the left peak corresponding to the lower critical temperature $T_{c}^2$. In Fig.~\ref{fig1}(c), $T_{c}^2$ has been extracted from the intersection of $U_m$ curves for various $L$, and in Fig.~\ref{fig1}(d), $T_{c}^1$ is quantified from the intersection of $U_4$. $T_{c}^1$ and $T_{c}^2$ for 9-state clock model are $\sim$ 0.9 and 0.33 respectively.

\begin{table}[!htbp]
\caption{Lower critical temperatures $T_{c}^2 (q)$ for the $d=2$ $q$-state clock model.} 
\label{table_Q}
\begin{center}
\begin{tabular}{lc}
\hline
\hline
$q$ & $T_{c}^2$ \\
\hline 
5 & 0.897 $\pm$ 0.001 \\
6 & 0.681 $\pm$ 0.001 \\
7 & 0.531 $\pm$ 0.006 \\
8 & 0.418 $\pm$ 0.001 \\
9 & 0.334 $\pm$ 0.001 \\
12 & 0.189 $\pm$ 0.002 \\
20 & 0.0695 $\pm$ 0.0003 \\
25 & 0.0448 $\pm$ 0.0002 \\
35 & 0.0235 $\pm$ 0.0006 \\
\hline
\hline
\end{tabular} 
\end{center}
\end{table}

\begin{figure}[!htbp]
\centering
\includegraphics[width=\columnwidth]{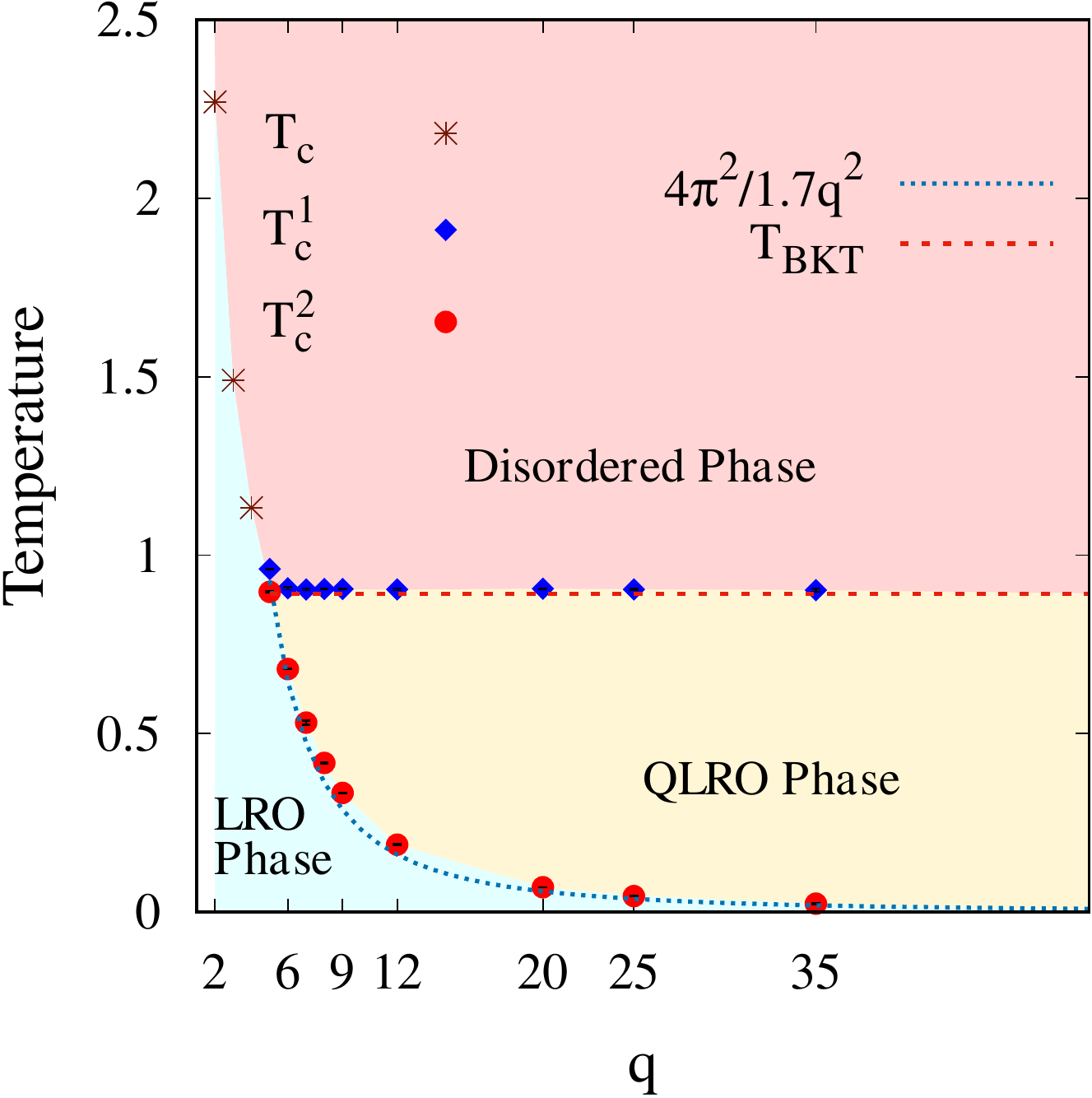}
\caption{(Color online) Phase diagram of the $q$-state clock model. Simulation data shows that $T_{c}^2$ (red solid circle) is a decreasing function of $q$ ($\forall$ $q$ $\geqslant$ 5) whereas $T_{c}^1$ (blue solid rhombus) remain constant at $\simeq$ 0.9. Figure also shows single phase transition for $q \leqslant 4$ at $T_c$ (maroon star). The errors (black error bars) are smaller than the corresponding symbol sizes. The two dashed lines fitted with $T_{c}^1$ and $T_{c}^2$ are the theoretical prediction of Berezinskii-Kosterlitz-Thouless transition temperature, $T_{BKT}$ $\simeq$ 0.892 and second transition temperatures of the $q$-state clock model, $T_{c}^2(q)$ = $4\pi^2/1.7q^2$, respectively.}
\label{fig2}
\end{figure}

The $q$-state clock model phase diagram in Fig.~\ref{fig2}, essentially spotlights the quantitative change of $T_{c}^1$ and $T_{c}^2$ with $q$. The plot clearly demonstrates that the phase transition for $q \leqslant 4$ is characterized by one transition temperature $T_c$, represented by maroon star (color online). It also comprehensively shows that $T_{c}^1$, represented by blue solid diamond (color online), approaches $\simeq$ 0.9 for $q \geqslant 6$, while $T_{c}^2$, represented by red solid circle (color online), is a decreasing function of $q$ and for large enough $q$ this transition eventually vanishes. Clearly, with increasing number of spin states the discrete clock model becomes identical with $d=2$ $XY$ model ($q$ $\rightarrow$ $\infty$) with only one phase transition occurring at the Kosterlitz-Thouless transition point $T_{KT}$ $\simeq$ 0.892. The dashed line fitted with $T_{c}^2$ is the analytical prediction of the lower transition temperature $T_{c}^2/J \simeq 4\pi^2/1.7q^2$ \cite{kadanoff77,wexler2006,nelson83}. $T_{c}^2$ as a function of $q$ is tabulated in Table \ref{table_Q}. We have also explicitly marked the three different phases, disordered (light pink, color online), QLRO (pale yellow, color online), and LRO (light sky blue, color online) in order to have a more clear understanding of the phase diagram. Our results agrees with the previously estimated values of $T_{c}^2$ obtained using different approaches \cite{brito2010,wexler2006}. We utilize this universal phase diagram for $q$-state clock model to characterize the kinetics of domain growth following temperature quenches in the LRO and QLRO regimes.

\subsection{Ordering dynamics in $q$-state clock model}

In this section, we present numerical results of phase ordering kinetics in $q$-state clock model for $q$ = 6, 9, 12, and 20. The simulations are carried out on a square lattice of size $1024^2$ having periodic boundary conditions in all directions. To emulate the homogeneous phase at high $T$, we assign random initial orientations to each spin according to Eq.~\eqref{theta} and rapidly quench the system from $T$ = $\infty$ to (a) $T$ = 0.1 in the LRO regime ($T<T_{c}^2$, see Table \ref{table_Q}) and (b) $T$ = 0.8 for 6-state clock model and $T$ = 0.5 or 0.6 for 9-, 12-, and 20-state clock models which are within the QLRO regime for the respective $q$ states ($T_{c}^2<T<T_{c}^1$). Using Metropolis algorithm, the system is updated up to $t$ = $10^6$ MCS. All statistical results presented in this section are averaged over 20 independent initial realizations. 

\begin{figure}[!htbp]
\centering
\includegraphics[width=\columnwidth]{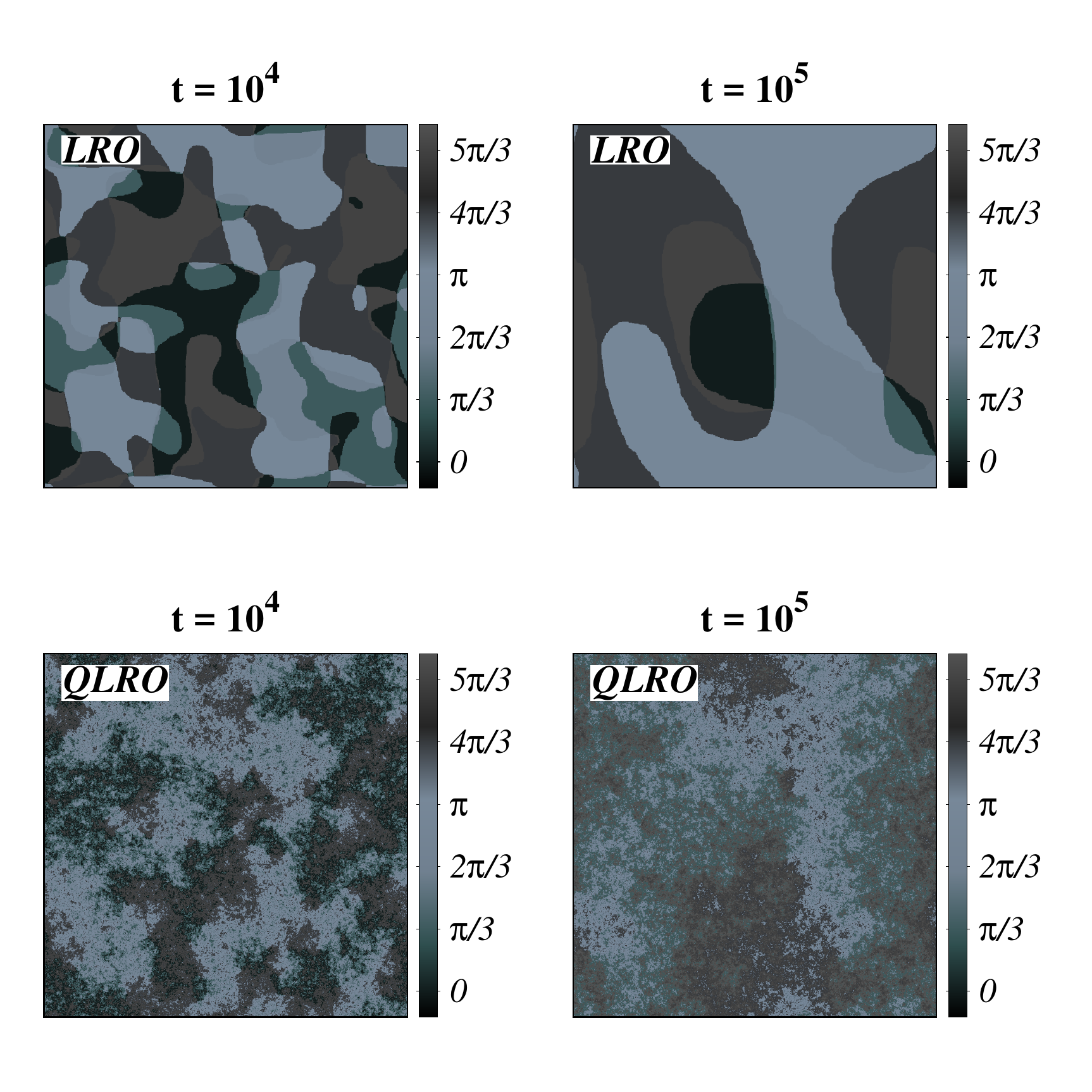}
\caption{(Color online) Time evolution snapshots of the 6-state clock model at $t=10^4$ MCS and $t=10^5$ MCS for a quench to the LRO (upper panel) and QLRO (lower panel) regimes are shown on square lattices of size $1024^2$. Upper panel shows domain evolution morphologies for a quench from $T$ = $\infty$ to $T$ = 0.1 (LRO, $T<T_{c}^2$) whereas domain morphologies for a quench to $T$ = 0.8 (QLRO, $T_{c}^2<T<T_{c}^1$) are shown in the lower panel. Grey color shades indicates different angles possible for the orientations of the clock spins according to Eq.~\eqref{theta}. The data clearly shows that quench to the LRO regime is characterized via well-defined compact domains whereas interpenetrating rough domain morphology is the salient feature of the QLRO regime.}
\label{fig3}
\end{figure}

Time evolution snapshots of domains for 6-state clock model at $t$ = $10^4$ MCS and $10^5$ MCS on a $1024^2$ lattice are shown in Fig.~\ref{fig3}. Upper panel represent domain evolution for a quench to the LRO regime at $T$ = 0.1 and in the lower panel quench is done to the QLRO regime at $T$ = 0.8. We see well-defined domains grow significantly at later times for the quench to the LRO regime, whereas, interpenetrating domains with rough interfaces is typically the signature in the QLRO regime. In the latter regime, no sharp domain boundaries between neighboring domains could be observed. The colorbar consists of six different shades in grey (color online) corresponding to the six possible orientations for the spin vectors of a 6-state clock model as per Eq.~\eqref{theta}.

\begin{figure}[!htbp]
\centering
\includegraphics[width=\columnwidth]{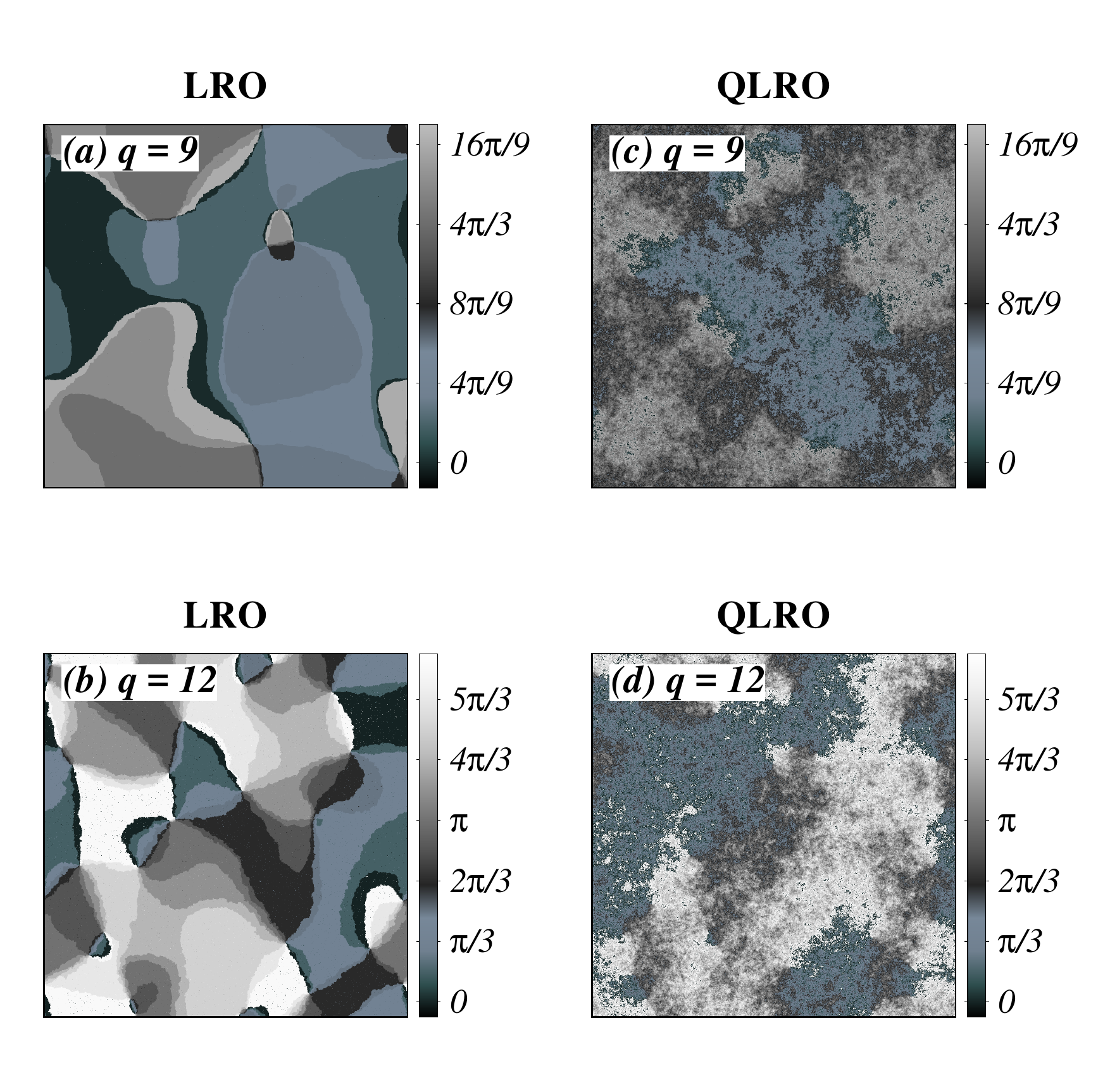}
\caption{(Color online) [(a)-(b)] Domain evolution snapshots of the (a) 9-state clock model and (b) 12-state clock model at $t$ = $10^5$ MCS after a quench from $T$ = $\infty$ to the LRO regime ($T$ = 0.1, $T<T_{c}^2$). The lattice size is $1024^2$. [(c)-(d)] Analogous to (a) and (b) but for a quench to the QLRO regime ($T$ = 0.5, $T_{c}^2<T<T_{c}^1$). Orientation of the clock spins according to state $q$ are represented by the different grey color shades. Due to higher degeneracy of the ground state, average domain size for $q=12$ is less than $q=9$. A similar comparison is not very obvious from the snapshots presented in (b) and (d).}
\label{fig4}
\end{figure}

Fig.~\ref{fig4} shows domain evolution snapshots for 9 and 12-state clock model at $t$ = $10^5$ MCS. Fig.~\ref{fig4}(a) and Fig.~\ref{fig4}(b) respectively show domain configurations for $q$ = 9 and 12 after a quench from T = $\infty$ to $T$ = 0.1 (LRO, $T<T_{c}^2$). QLRO domain configurations after a rapid quench to $T$ = 0.5 for the above mentioned $q$ states are shown in Fig.~\ref{fig4}(c) and Fig.~\ref{fig4}(d) respectively. Lattice size is $1024^2$. As we see in the upper panel of Fig.~\ref{fig3}, Fig.~\ref{fig4}(a) and Fig.~\ref{fig4}(b) show distinct domain structures whereas in Fig.~\ref{fig4}(c) and Fig.~\ref{fig4}(d), domains are interpenetrating and lack compactness. Grey shades (color online) in the colorbars represent different possible angles of the spin vectors at each lattice point.

A $q$-state clock model can have $q$ possible ground states and the interface between two neighboring domains eventually lead to $^qC_2 = \frac{q(q-1)}{2}$ different choices. Moreover, in $d=2$ and $q\geqslant 3$, three or more different domains would meet at a point [see Fig.~\ref{fig4}(a) and Fig.~\ref{fig4}(b)] and for such a system both interfaces and point defects (vortices and anti-vortices) are present as topological disorders \cite{Bray94}. Vortex and antivortex are identified by the net change in spin orientations surrounding the point defects: if the change is $2\pi$, it is a vortex and if the change is $-2\pi$, the defect is an anti-vortex. Initially the system coarsens via merging of domain walls and annihilation of vortices by oppositely charged anti-vortices; nevertheless, a close observation at Fig.~\ref{fig3} snapshots in the LRO regime reveal that asymptotically merging of the domain interfaces becomes a dominant mechanism in the growth process. As the system approaches toward the equilibrium, energetically expensive interfaces and point defects are rapidly eliminated and at a later stage, a few thermally exited interfaces and defects remain in the system. The colormaps in Fig.~\ref{fig4}(a) and Fig.~\ref{fig4}(b) also indicate that when the average angular difference between the adjacent domains is small, energy cost to create an interface is minimum and such configurations are easy to find. Vortices are clearly observed in the domain morphologies when the quench is done in the LRO regime. There one can notice points encircling by a number of different phases or domains and the phases are arranged in such a way that the average angular difference of spin orientations of the two adjacent domains are always minimum. We have also noticed that at very late time of the growth process, system is completely free of point defects whereas interfaces are still present. Domains in Fig.~\ref{fig4}(b) are more crowded (single spin domains) than of Fig.~\ref{fig4}(a) as the ground state of $q=12$ is more degenerate than $q=9$. 

\begin{figure}[!htbp]
\centering
\includegraphics[width=\columnwidth]{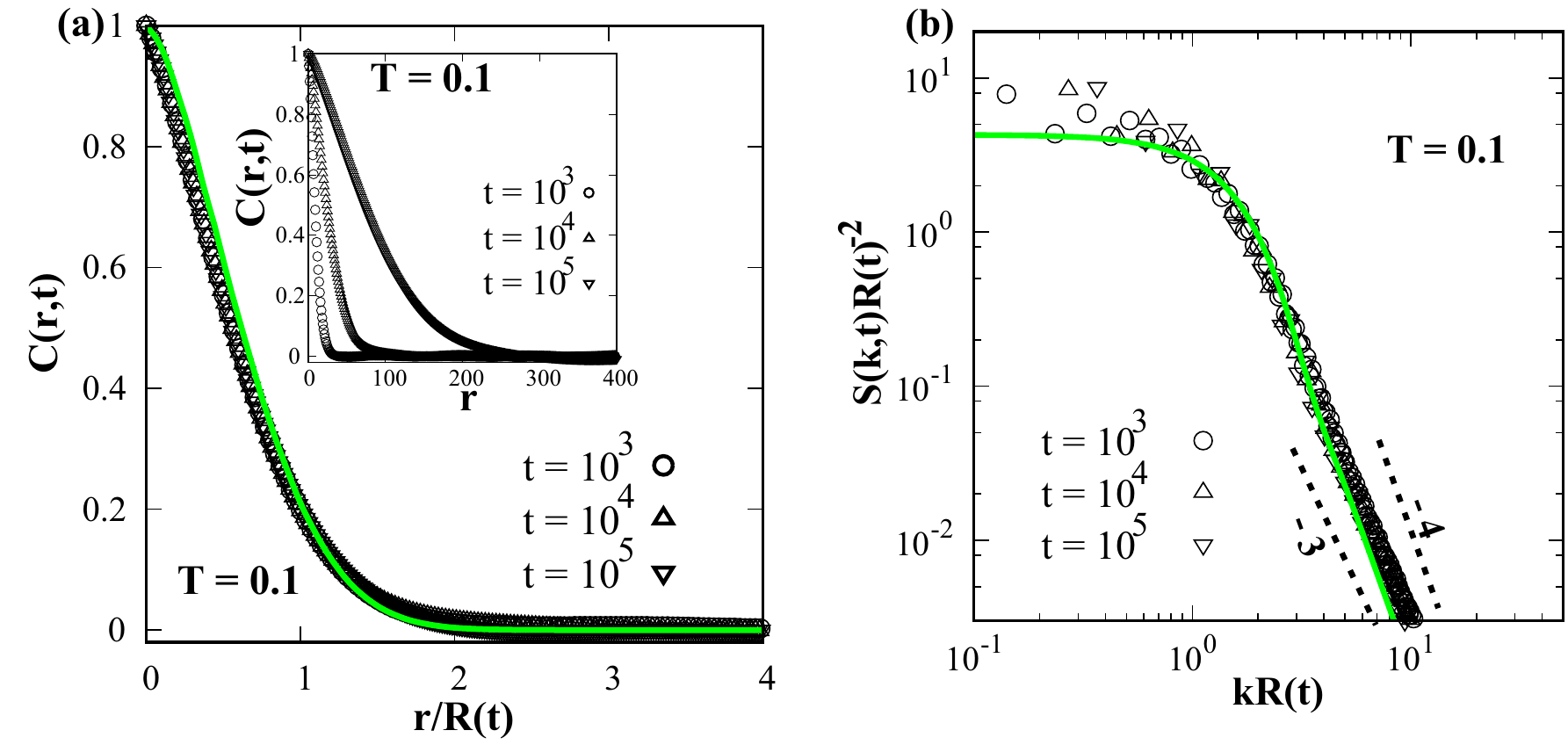}
\caption{(Color online) Dynamical scaling of 9-state clock model for a quench to the LRO regime. (a) Scaling plot of the correlation function $C(r,t)$ versus $r/R(t)$ after a quench from $T = \infty$ to $T = 0.1$ ($T<T_{c}^2$). The inset shows the unscaled data. (b) Scaled structure factor (on a log-log scale), $S(k,t)R(t)^{-2}$ versus $kR(t)$. The Bray-Puri-Toyoki (BPT) function in Eq.~\eqref{bpt} for $n=2$, and its Fourier transform are shown as the (green) solid curves in (a) and (b), respectively. Slope of the large-$k$ tail of the structure factor scaling function found to lie between $-3$ and $-4$ (-3.27 precisely). A $-3$ slope of the structure factor tail signifies the Porod's decay, $S(k,t) \sim k^{-(d+1)}$, whereas, $-4$ slope denotes the generalized Porod's law: $S(k,t)$ $\sim$ $k^{-(d+n)}$ for $d$ = $n$ = 2.}
\label{fig5}
\end{figure}

Our data, in Fig.~\ref{fig5} present the dynamical scaling of the correlation function and structure factor for 9-state clock model when the system is quenched to $T$ = 0.1 ($T<T_{c}^2$) in the LRO regime. Note that, a similar dynamical scaling in correlation function and structure factor holds good for $q$ = 6 and 12 as well (data not shown here). Data collapse in Fig.~\ref{fig5}(a) confirms the dynamical scaling of the correlation function $C(r,t)$ with $r/R$; whereas structure factor $S(k,t)$ suggests a scaling form $S(k,t)R(t)^{-2}$ with $kR(t)$ as shown in Fig.~\ref{fig5}(b). The structure factor $S(k,t)$ is essentially as the Fourier transform of the equal time correlation function $C(r,t)$. Physically, dynamical scaling of the correlation function signifies that the domain morphologies are equivalent and independent of time when characteristic lengths are scaled by average domain size $R(t)$ at time $t$. The solid curve in Fig.~\ref{fig4}(a) is the Bray-Puri-Toyoki (BPT) scaling function~\cite{BP91,toyoki92} for $n=2$. The BPT scaling function $f(x)$ is the generalization of the Ohta-Jasnow-Kawasaki function \cite{OJK} for $n$-component Time-Dependent Ginzburg-Landau (TDGL) equation (with $n \leqslant d$) and has the form:
\begin{equation}
\label{bpt}
f(r/R)=\frac{n\gamma}{2\pi}\left[B\left(\frac{n+1}{2},\frac{1}{2}\right)\right]^2 F\left(\frac{1}{2},\frac{1}{2};\frac{n+2}{2};\gamma^2\right),
\end{equation}
where $\gamma=\exp(-r^2/R^2)$, $B(x,y)\equiv \Gamma(x)\Gamma(y)/\Gamma(x+y)$ is the Euler's beta function, and $F(a,b;c;z)$ is the hypergeometric function $_2F_1$. In Fig.~\ref{fig5}(b), the solid curve is the Fourier transform of the BPT function and the large-$k$ behavior of the structure factor tail generate a slope -3.27 (in log-log plot) lying between -4 and -3. A -4 slope of the structure factor tail corresponds to the `generalized Porod's law': $S(k,t)\sim k^{-(d+n)}$ for $d=2, n=2$ and a growth process typically driven by point defects, such as vortices and antivortices (this is because the defect core of an $n$-component model in $d$ dimensions will have a surface of dimension $d-n$ \cite{Bray94} and for $d=n=2$, these defects are points) whereas slope -3 denotes the `Porod's decay': $S(k,t)\sim k^{-(d+1)}$ for compact domains surrounded by sharp domain interfaces. Structure factor tail with slope between -4 and -3 physically implies a growth process involving both types of topological defects which could be explained from the corresponding length scale data in Fig.~\ref{fig7}. 

\begin{figure}[!htbp]
\centering
\includegraphics[width=\columnwidth]{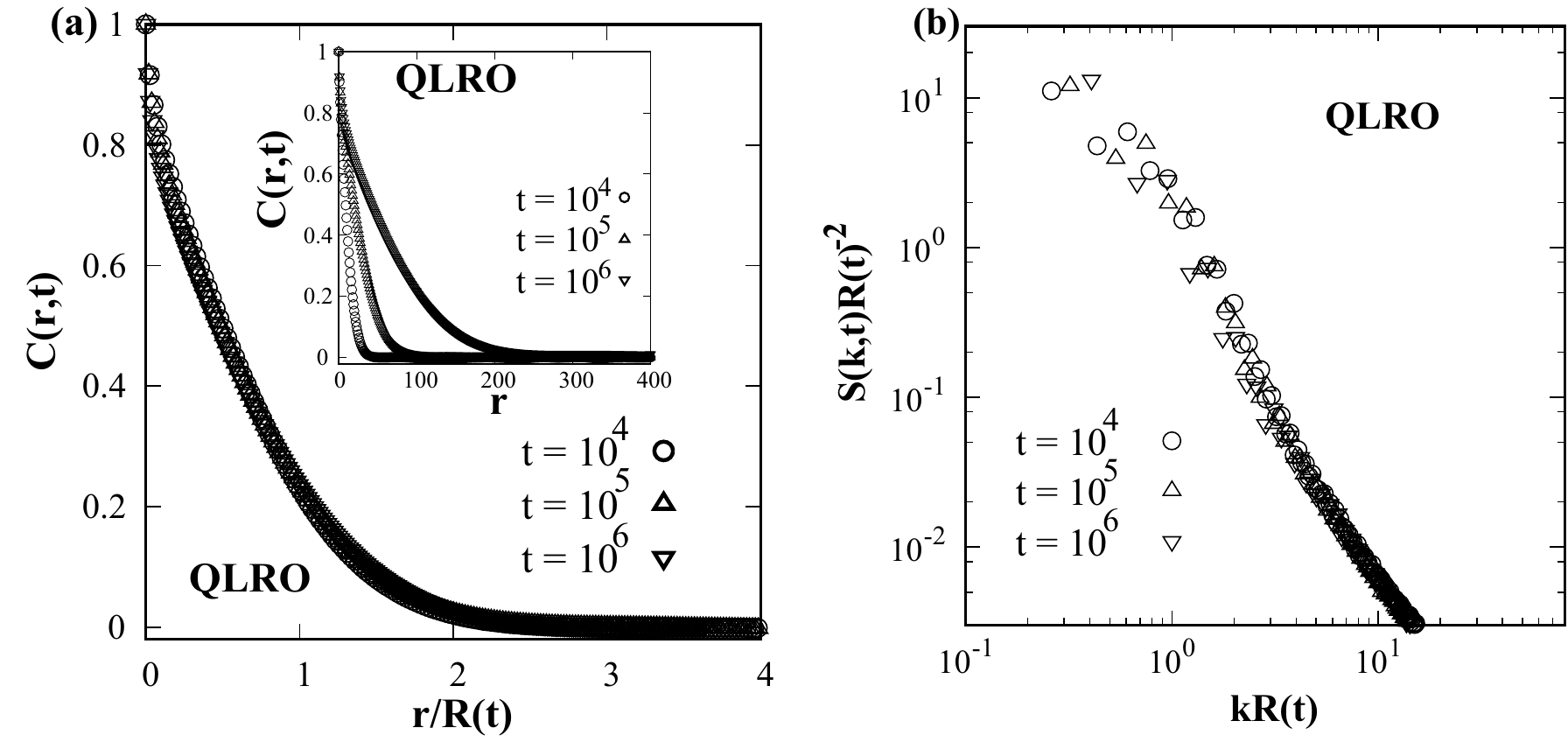}
\caption{(a) Scaled correlation function, $C(r,t)$ versus $r/R$, for 9-state clock model after a quench from $T = \infty$ to the QLRO regime ($T_{c}^2<T<T_{c}^1$) along with the unscaled data (inset) at $t$ = $10^4$, $10^5$, and $10^6$ MCS. (b) Scaled structure factor in log-log plot. Data shows that dynamical scaling is valid for a quench between $T_{c}^1$ and $T_{c}^2$ but the same is not validated by the BPT function (not shown) as in Fig.~\ref{fig5}.}
\label{fig6}
\end{figure}

\begin{figure}[!htbp]
\centering
\includegraphics[width=\columnwidth]{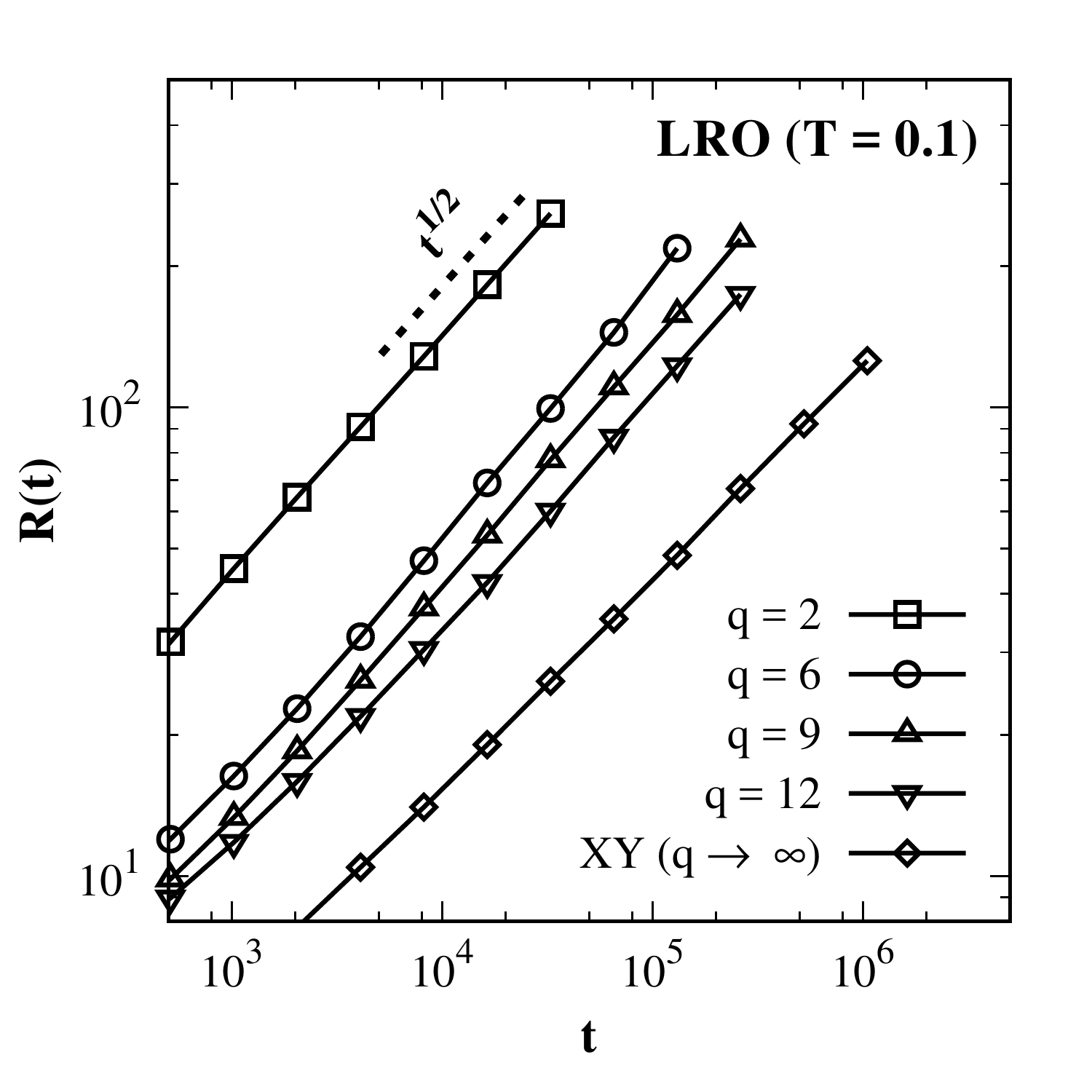}
\caption{$R(t)$ versus $t$ (on a log-log scale) for various values of $q$ following a quench from $T = \infty$ to the LRO regime ($T = 0.1$). The dashed line indicates the growth law $R(t)$ $\sim$ $t^{1/2}$ and is included as a guide to the eye. The length scale data of the $q$-state clock model are further compared with the length scale data of the $d=2$ $XY$ model for which the extracted slope is $<0.5$ as $XY$ model exhibit a growth law: $R(t)$ $\sim$ $(t/\ln t)^{1/2}$.}
\label{fig7}
\end{figure}

Next we examine the scaling properties of the correlation function and structure factor for a quench in the QLRO regime, $T_{c}^2<T<T_{c}^1$. Data shown here are for a 9-state clock model; qualitative nature of the data for $q$ = 12 and 20 are similar (not shown). We plot $C(r,t)$ versus $r/R$ along with the unscaled data in the inset for $t$ = $10^4$ MCS, $10^5$ MCS, and $10^6$ MCS in Fig.~\ref{fig6}(a) and observe nice data-collapse on a single master curve implying that the domain morphologies are time invariant. Fig.~\ref{fig6}(b) shows the scaled structure factor. The asymptotic regime of the structure factor scaling function or the large-$k$ tail of the scaled structure factor in Fig.~\ref{fig6}(b) is characterized by $S(k,t)\sim k^{-1.9}$, where the non-integer slope -1.9 is indicative of the interpenetrating fractal architecture of domains or systems with rough morphologies [see Fig.~\ref{fig4}(b)-(d)] where scattering happens from fractal interfaces. These data clearly indicate a non-Porod behavior for a quench in the QLRO regime \cite{corberi2006}. Non-integer decay exponent of the structure factor tail has also been reported for other systems representing ground-state morphologies of a dilute anti-ferromagnets, droplet-in-droplet morphologies of a double-phase-separating mixtures, ground-state morphologies of the $d=2,3$ random-field Ising Model \cite{SP2014,MK2014,VB2016} etc., where the respective non-integer exponents are found to exhibit $non-Porod$ behavior associated with scattering from the rough domain morphologies.

Subsequently, we investigate the time dependence of the average domain length $R(t)$.

A rapid quench from a high temperature homogeneous phase to a temperature below the critical temperature makes a system thermodynamically unstable and subsequent evolution toward the new equilibrium state emerges a length-scale $R(t)$ corresponding to the preferred phase. $R(t)$ is the characteristic length scale of domains that grow with $t$. In systems with sharp domain interfaces, the driving force for late stage domain growth is the domain wall curvature, as system dissipates energy by contracting the total surface area. The relation between interface motion and local curvature as per Allan-Cahn \cite{CA} equation is $v=-K$, where $v \sim dR/dt$ is the interface velocity and $K \sim 1/R$ is the domain wall curvature. For a nonconserved system, the equation for curvature-driven growth reads \cite{Bray94} \cite{PW}:
\begin{equation}
\label{LCA}
\frac{dR}{dt} = \frac{a(R,T)}{R}
\end{equation} 
where $a(R,T)$ is the diffusion constant and depends on length scale $R$ and temperature $T$. For pure systems, $a(R,T)$ is invariant of domain scale and temperature, $i.e.$, $a(R,T)$ = constant. Now, equating and integrating \eqref{LCA} we get the domain growth law for curvature driven growth as $R(t)\sim t^{\frac{1}{2}}$, the Lifshitz-Cahn-Allen (LCA) growth law. The LCA growth law has also been widely reported as the governing domain growth law for $q$-state clock model in the literature \cite{Bray94,PW}. Nevertheless, as we see in the domain evolution snapshots presented in Fig.~\ref{fig4}(a) and Fig.~\ref{fig4}(c) that sharp domain walls can only be seen when the temperature quench is done in the LRO regime $i.e.$ $T<T_{c}^2$. Quench in the QLRO regime, where $T_{c}^2<T<T_{c}^1$, led to a coarsening via interaction of rough domain interfaces. Thus, we argue that for domain coarsening in $q$-state clock model, LCA growth law is only valid for a quench into the LRO regime. Fig.~\ref{fig7} shows the plots of $R(t)$ versus $t$ on a log-log scale for $q$ = 2-, 6-, 9-, and 12-state clock model and $d=2$ $XY$ model with quench temperature $T$ = 0.1. $R(t)$ is measured from the correlation function $C(r,t)$ when it falls to 0.3 of its maximum value. Fitting a straight line with the simulation data we have extracted the asymptotic growth exponent $n$ $\simeq$ 0.5 for $q$ = 2, 6, 9, and 12, indicated by the dashed line placed as a guide to the eye. The exponent is less than 0.5 for the $XY$ model as ordering kinetics in $d=2$ $XY$ model follows $R(t)\sim (t/\log t)^{\frac{1}{2}}$ growth law. A close observation of the $R(t)$ versus $t$ plot for $q$ = 6, 9, and 12 reveals a crossover from a preasymptotic domain growth with growth exponent $<0.5$ to an asymptotic growth with growth exponent $\simeq 0.5$. This behavior is possibly related to a relatively slow growth process in the preasymptotic regime driven by merging domain walls and annihilating point defects that switches to a faster growth regime driven majorly by merging domain interfaces leading to a $t^{1/2}$ growth law in the asymptotic limit. This signature can further be seen in Fig.~\ref{fig5}(b) where the tail of the structure factor scaling function decays with an effective exponent -3.17 signifying contribution from both types of defects (domain walls and point defects) in the growth process in the initial stage and the late time behavior is mainly controlled by the sharp domain interfaces.

\begin{figure}[!htbp]
\centering
\includegraphics[width=\columnwidth]{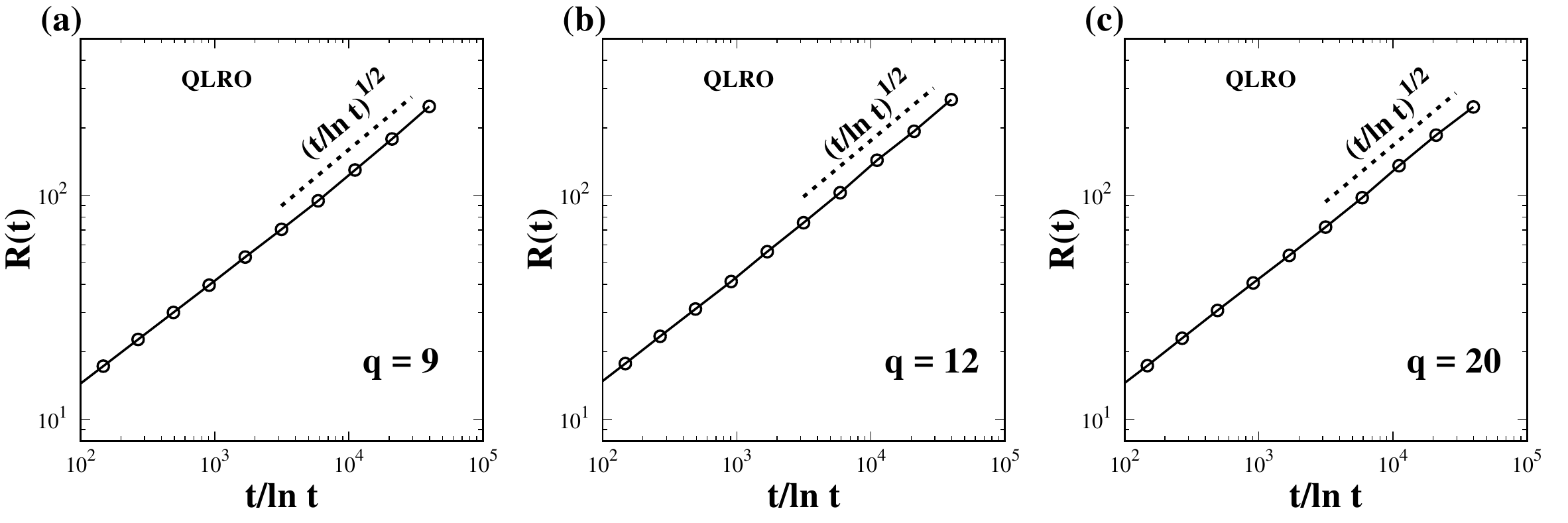}
\caption{Length scale data for $q$-state clock model for a quench from $T = \infty$ to the QLRO regime ($T$ = 0.5, $T_{c}^2<T<T_{c}^1$) of the respective $q$-state clock models. Plot of $R(t)$ versus $t/\ln t$ (on a log-log scale) are shown for (a) $q$ = 9, (b) $q$ = 12, and (c) $q$ = 20 state clock models. The $d=2$ $XY$ growth law $R(t)$ $\sim$ $(t/\ln t)^{1/2}$ is indicated by the dashed line which is of slope 0.5. Extracted exponents for the 9-state, 12-state, and 20-state clock models are 0.501$\pm$0.002, 0.504$\pm$0.008, and 0.495$\pm$0.009 respectively.}
\label{fig8}
\end{figure}

In Fig.~\ref{fig8}, we demonstrate the possible growth law for $q$-state clock model for a quench in the QLRO regime. $R(t)$ versus $t$ plots for $q$-state clock model with quenched temperatures between $T_{c}^1$ and $T_{c}^2$ yield power-law growth with exponents much less than 0.5 for $q$ = 6, 9, 12, and 20 (data not shown). For $q$ = 6, the growth law reads $R(t) \sim t^{0.38}$ at quenched $T$ = 0.8 is in good agreement with an earlier study \cite{corberi2006} where $R(t) \sim t^{0.35}$ for a quench to $T$ = 0.76. This slow domain growth law is a natural consequence of interpenetrating domains or in other words, the growth process is governed by fractal interfaces. It is imperative to understand the domain growth kinetics in clock model from our existing knowledge of coarsening dynamics in the $d=2$ $XY$ model. Pure $XY$ model ($n=2$) in $d=2$ ($q$ $\rightarrow$ $\infty$) has two phases, a high temperature disordered phase and a low temperature QLRO phase (below Kosterlitz-Thouless transition temperature) and is completely devoid of any LRO phase. Domain growth kinetics for a temperature quench to the QLRO phase in $d=2$ $XY$ model or in general a model with $d=n=2$, is characterized by the $R(t) \sim (t/\ln t)^{\frac{1}{2}}$ \cite{Bray94,BR94}. Now for $d=n=2$, by definition of defect dimension, this growth should be controlled by dimensionless point defects which are vortices and antivortices. The QLRO phase of $q$-state clock model is almost devoid of well-defined domain interfaces and coarsening can be explained only via the annihilation of vortex-antivortex pairs. In our data in Fig.~\ref{fig8}[(a)-(c)], we have shown that a logarithmic correction of $t$ yields growth exponents $n$ $\simeq$ 0.5 for $q$ = 9, 12, and 20 in the asymptotic time limit. The respective growth exponents which we measure during $t\in [10^4-10^5]$ for $q$ = 9, 12, and 20 by fitting a straight line with the simulation data are 0.501$\pm$0.002, 0.504$\pm$0.008, and 0.495$\pm$0.009. For a quench to the QLRO regime, our numerical results demonstrate that coarsening dynamics in this regime is driven by the elimination of vortex-antivortex pairs and $R(t) \sim (t/\ln t)^{\frac{1}{2}}$ is the asymptotic growth. Since $T_c^2$ is a decreasing function of $q$, at large $q$ (as well as $q$ $\rightarrow$ $\infty$, $XY$ model) the same growth law persists across the temperatures below $T_c^1$ dominated by a QLRO phase.

\section{Summary and Discussion} 
\label{Summary}

In this paper we have undertaken a numerical investigation of the non-equilibrium domain growth kinetics of the $q$-state clock model for a quench to the LRO and QLRO phase via comprehensive Monte Carlo simulation. We first confirm the existence of two distinct transition temperatures for a finite $q$-state clock model with $q \geqslant 5$ \cite{kadanoff77,elitzur79,domany80,cardy80,tobochnik82,baek2010,brito2010} and then quantified the transition temperatures. Transition from the disordered to QLRO phase at $T_{c}^1$ is quantified from the temperature dependence of Binder cumulant $U_4$ [see Eq.~\eqref{BC}] and the lower transition temperature $T_{c}^2$ which characterizes the transition from QLRO phase to LRO phase, is measured from the temperature dependence of another cumulant $U_m$ [see Eq.~\eqref{um}]. 

We have investigated the coarsening dynamics of $q$-state clock model by quenching an initially prepared homogeneous system at $T\rightarrow \infty$ to $T<T_{c}^2$, where the equilibrium phase is LRO and $T_{c}^2<T<T_{c}^1$, where the equilibrium phase is QLRO. The domain morphologies for the corresponding quenches are investigated from the behavior of the equal time spatial correlation function $C(r,t)$ and its Fourier transform, the structure factor $S(k,t)$. 

The quench to the LRO phase ($T<T_{c}^2$) is characterized by well-defined domain boundary with vortices or antivortices at the meeting point of three or more sharp domain interfaces. The scaling form of $C(r,t)$ and $S(k,t)$ are time-invariant and the large-$k$ tail of the structure factor scaling function yields a slope falling between the slope of Porod decay and generalized Porod law. The growth law in this regime exhibit a crossover from a slow power-law growth in the preasymptotic regime where coarsening is governed by the annealing of both vortices and interfaces to a LCA power-law growth at the asymptotic limit where growth is characterized mainly by the merging of sharp domain interfaces. 

The characteristic of domains for a quench to the QLRO phase ($T_{c}^2<T<T_{c}^1$) can be well described by the interpenetrating, rough domain morphology with well-defined point defects. Although the domain growth in this regime satisfies the power-law $R(t) \sim t^n$, $n$ turns out to be $<0.5$. This slow non-equilibrium dynamics is a consequence of fractal interfaces between neighboring domains. Our length scale data for $q$ = 9, 12, and 20 convincingly establish that coarsening in the QLRO phase is related to the growth law for $d=n=2$ systems akin to the $XY$ model, where the growth law is: $R(t) \sim (t/\ln t)^{\frac{1}{2}}$. Since the QLRO phase of the $q$-state clock model does not possess well-defined sharp domains, coarsening dynamics in this regime is driven via the annihilation of point defects, viz., vortices and antivortices.

In conclusion, our study provides a detail analysis of domain growth kinetics of finite $q$-state clock models in both LRO and QLRO regimes of the phase diagram. The asymptotic growth kinetics in these regimes is driven by two different mechanisms, which are entangled in the early time regimes. Thus, an interesting extension of this study would be to explore the phase ordering kinetics in presence of disorder. 


\section{Acknowledgements}

S.C. thanks CSIR, India, for support through Grant No. 09/080(0897)/2013-EMR-I. R.P. thanks CSIR, India, for support through Grant No. 03(1414)/17/EMR-II. S.P. is grateful to the Department of Science and Technology, India for funding through a J.C. Bose fellowship.


\end{document}